\documentclass[12pt,preprint,iop,apj]{emulateapj}

\usepackage{graphicx}
\usepackage{txfonts}
\newcommand{\va}{v_{\mathrm{A}}}
\newcommand{\vai}{v_{\mathrm{Ai}}}
\newcommand{\vae}{v_{\mathrm{Ae}}}

\newcommand{\ta}{\tau_{\mathrm{A}}}

\newcommand{\ui}{U_{\mathrm{i}}}
\newcommand{\ue}{U_{\mathrm{e}}}

\newcommand{\pd}{\partial}

\newcommand{\vkf}{{v_{\rm kf}^{\pm}}}
\newcommand{\vke}{v_{\rm KE}}
\newcommand{\vcm}{v_{\rm cm}}

\newcommand{\vk}{v_{\rm k}}

\newcommand{\rhoi}{\rho_{\rm i}}
\newcommand{\rhoe}{\rho_{\rm e}}

\newcommand{\ld}{L_{\rm D}}

\begin{document}

	\title{SPATIAL DAMPING OF PROPAGATING KINK WAVES DUE TO RESONANT ABSORPTION: EFFECT OF BACKGROUND FLOW}

	\shorttitle{RESONANT KINK WAVES WITH FLOW}

   \author{R. Soler$^1$, J. Terradas$^2$, and M. Goossens$^1$}
   \affil{$^1$Centre for Plasma Astrophysics, Katholieke Universiteit Leuven,
              Celestijnenlaan 200B, 3001 Leuven, Belgium}
              \email{roberto.soler@wis.kuleuven.be}

 \affil{$^2$Departament de F\'isica, Universitat de les Illes Balears,
              E-07122, Palma de Mallorca, Spain}

  \begin{abstract}
  
Observations show the ubiquitous presence of propagating magnetohydrodynamic (MHD) kink waves in the solar atmosphere. Waves and flows are often observed simultaneously. Due to plasma inhomogeneity in the perpendicular direction to the magnetic field, kink waves are spatially damped by resonant absorption. The presence of flow may affect the wave spatial damping. Here, we investigate the effect of longitudinal background flow on the propagation and spatial damping of resonant kink waves in transversely nonuniform magnetic flux tubes. We combine approximate analytical theory with numerical investigation. The analytical theory uses the thin tube (TT) and thin boundary (TB) approximations to obtain expressions for the wavelength and the damping length. Numerically, we verify the previously obtained analytical expressions by means of the full solution of the resistive MHD eigenvalue problem beyond the TT and TB approximations. We find that the backward and forward propagating waves have different wavelengths and are damped on length scales that are inversely proportional to the frequency as in the static case. However, the factor of proportionality depends on the characteristics of the flow, so that the damping length differs from its static analogue. For slow, sub-Alfv\'enic flows the backward propagating wave gets damped on a shorter length scale than in the absence of flow, while for the forward propagating wave the damping length is longer. The different properties of the waves depending on their direction of propagation with respect to the background flow may be detected by the observations and may be relevant for seismological applications.

  \end{abstract}



\section{Introduction}

The presence of ubiquitous small-amplitude propagating magnetohydrodynamic (MHD) waves in magnetic waveguides of the solar corona was first observed with the Coronal Multi-channel Polarimeter (CoMP) \citep[see][]{tomczyk07,tomczyk09}. The observations have inspired a number of recent theoretical works in which the properties of propagating waves are studied. Based on MHD wave theory \citep[e.g.,][]{erdelyi2007,tom08,tom08b,VTG} the observations have been interpreted as propagating kink waves, i.e., transverse MHD waves with mixed fast and Alfv\'enic properties whose dominating restoring force is magnetic tension \citep[see, e.g.,][]{edwinroberts,goossens2009}. In particular, \citet{tom08b} showed that field-aligned density enhancements in the corona act as natural waveguides for MHD waves. Apart from the CoMP observations in coronal loops, propagating kink waves have also been observed in chromospheric spicules \citep[e.g.,][]{depontieu07,he1,he2} and in thin threads of solar prominences \citep[e.g.,][]{lin07,lin09}. In the same manner as standing kink MHD waves are damped in time by resonant absorption \citep[see, e.g.,][]{goossens2002,rudermanroberts}, propagating kink MHD waves are damped in space due to naturally occurring plasma inhomogeneity in the direction transverse to the the magnetic field \citep{TGV}. \citet{TGV}, hereafter TGV, obtained the important result that the damping length by resonant absorption is inversely proportional to the wave frequency. This means that high-frequency waves are damped in shorter length scales than low-frequency waves. \citet{VTG} showed that the analytical theory of propagating resonant kink waves developed by TGV is fully consistent with the CoMP observations. For the time-dependent, driven problem \citet{pascoe} studied numerically the spatial damping of resonant kink waves and obtained equivalent results to those analytically predicted by TGV. Importantly, \citet{solerspatial} have shown that the results of TGV still hold when the plasma is partially ionized, so that the theory can be applied to kink waves propagating in the chromosphere and in prominences as well. Recently, \citet{stratified} extended the results of TGV by taking into account for the first time density variation both transversely and along the magnetic field direction.  

Field-aligned flows are also ubiquitous in magnetic structures in the solar atmosphere \citep[see the observational reports by, e.g.,][]{brekke,zirker,winebarger01,winebarger02,okamoto,chae2008,ofmanwang}. Typically, the observed flow velocities are smaller than 10\% of the plasma Alfv\'en speed. Faster flows of the order of the Alfv\'en speed are much less frequent and are related to energetic events as, e.g., flares and coronal mass ejections \citep[see, e.g,][]{innes}. The investigation of the effect of flow on the properties of the waves is therefore of evident interest. For example, the influence of flow and implications for MHD seismology of standing kink waves in coronal loops have been recently discussed by \citet{rudermanflow} and \citet{terradasletterflow}. In the case of prominences, \citet{solerthread} investigated standing kink waves in coronal flux tubes partially filled with flowing threads of prominence material. Nevertheless, none of these works took damping into account. Also for standing waves, \citet{terradasflow} studied temporal damping by resonant absorption in the presence of flow and found corrections to the damping time due to the flow with respect to the static case \citep{goossens2002}. However, in the case of spatial damping of propagating kink waves the effect of flow has not been investigated. TGV did not include flow in their study. To our knowledge there is no work in the literature that studies in detail the influence of flow on resonantly damped propagating waves in solar magnetic waveguides. Existing investigations of surface waves in the solar wind as, e.g., \citet{evans} did not perform a rigorous treatment of the process of resonant absorption and considered very simplified expressions for the damping length of the waves. Thus, a detailed investigation of propagating resonant MHD waves in the presence of flows in needed.

Here we investigate the effect of flow on the spatial damping of resonant kink waves in transversely nonuniform solar waveguides. We attack the problem both analytically and numerically. The analytical theory uses the thin tube (TT) and thin boundary (TB) approximations to obtain expressions for the wavelength and the damping length. We determine the influence of flow and compare our expressions with those obtained by TGV in the static case. Later, we use numerical methods to study the propagation and spatial damping of kink waves beyond the TT and TB approximations. The full numerical computations enable us to test the validity of the analytical expressions. Finally, we discuss the implications of our results for MHD seismology.

\section{Model and governing equations}
\label{sec:model}

The equilibrium configuration is a straight cylindrical magnetic flux tube of
radius $R$ embedded in a magnetized plasma environment. For convenience, we use
cylindrical coordinates, namely $r$, $\varphi$, and $z$ for the radial, azimuthal,
and longitudinal coordinates, respectively. The $z$-direction is set along the axis of the cylinder. 
We use the $\beta = 0$ approximation, where $\beta$ is the ratio of gas pressure to magnetic pressure.  The $\beta = 0$ approximation enables us to arbitrarily choose the density profile. In what follows, subscripts $\rm i$ and $\rm e$ refer to the internal and external plasmas, respectively. For example, we denote by $\rhoi$ and
$\rhoe$ the internal and external densities, respectively. Both of these quantities are
constants. There is a nonuniform transitional layer in the transverse
direction that continuously connects the internal density to the external density.
The layer has a thickness $l$ and covers the interval $R - l/2 \leq r \leq R
+ l/2$. The equilibrium magnetic field is straight, ${\bf B} = B
\hat{e}_z$, with $B$ constant. We assume a background flow along the magnetic field direction, ${\bf U} = U
\hat{e}_z$. We denote by $\ui$ and $\ue$ the internal and external flow velocities, respectively, which are constants. We take $\ui$ and $\ue$ as positive quantities. As for the density, we allow the flow velocity to change continuously in the radial direction from its internal to its external values within a transitional layer. The transitional layer for the flow velocity extends in the interval $R - l^\star/2 \leq r \leq R
+ l^\star/2$, with $l^\star$ the thickness of the transition.

Linear, ideal MHD waves propagating in our model are governed by the following set of equations,
\begin{eqnarray} 
\rho \left( \frac{\pd {\bf v}}{\pd t} + {\bf U} \cdot \nabla {\bf v} + {\bf v} \cdot \nabla {\bf U} \right) &=& \frac{1}{\mu} \left( \nabla
\times {\bf b} \right) \times {\bf B}, \label{eq:b1} \\ 
\frac{\pd {\bf b}}{\pd t}   -\nabla \times \left( {\bf U} \times {\bf b} \right) &=&
\nabla \times \left( {\bf v} \times {\bf B} \right), \label{eq:b2} 
\end{eqnarray}
where $\rho$ is the plasma density, ${\bf v}$ is the velocity perturbation, ${\bf b}$ is the magnetic field perturbation, and $\mu$ is the magnetic permittivity. As the equilibrium is uniform in both $\varphi$- and $z$-directions and we consider waves propagating along the tube with a fixed frequency, we write all
perturbations proportional to $\exp \left( i m \varphi + i k_z z - i \omega t \right)$, where
$m$ is the azimuthal wavenumber, $k_z$ is the longitudinal wavenumber, and $\omega$ is the wave angular frequency. Due to the presence of
a transverse inhomogeneous transitional layer, wave modes with $m \neq 0$ are
spatially damped due to resonant absorption. Here we are interested in kink waves, which are described by $m=1$. Kink waves have mixed Alfv\'enic and fast MHD properties. They are the only wave modes that can displace the magnetic cylinder axis and so produce transverse motions of the whole flux tube \citep[see, e.g.,][]{edwinroberts,goossens2009}. As a result of the process of resonant absorption, transverse kink motions of the flux tube are damped and azimuthal motions within the transitional layer are amplified as the wave propagates along the magnetic cylinder. 

For real $\omega$, resonant damping causes $k_z$ to be complex, $k_z = k_{z \rm R} + i k_{z \rm I}$, with $k_{z \rm R}$ and $k_{z \rm I}$ the real and imaginary parts of $k_z$, respectively. For fixed and positive $\omega$, the direction of wave propagation is determined by the sign of $k_{z \rm R}$. For $k_{z \rm R} > 0$ the wave propagates towards the positive $z$-direction (forward waves), whereas for $k_{z \rm R} < 0$ the wave propagates towards the negative $z$-direction (backward waves). In the absence of flow both directions of propagation are equivalent. In the presence of flow forward and backward waves have not the same properties and both directions of propagation must be taken into account \citep[see, e.g.,][]{nakaroberts,terra,solerflow,vashe}. Regarding the imaginary part of $k_z$, resonant absorption spatially damps the wave, so we expect $k_{z \rm I} > 0$. However, strong flows may trigger the Kelvin-Helmholtz Instability (KHI) \citep[see, e.g.,][]{chandra,drazin}, causing modes to be amplified in $z$, i.e., $k_{z \rm I} < 0$. From the real and imaginary parts of $k_z$ we compute the wavelength, $\lambda$, and the damping length, $\ld$, as 
\begin{equation}
 \lambda = \frac{2 \pi}{k_{z \rm R} }, \qquad \ld = \frac{1}{k_{z \rm I}}.
\end{equation}

\section{Analytical investigation}
\label{sec:analytics}

To study analytically the effect of flow on the resonantly damped kink waves we use the TT and TB approximations. In the TT approximation we restrict ourselves to waves with $\lambda / R \gg 1$. In terms of frequency, the TT approximation is equivalent to the low-frequency approximation, i.e., $\omega \ta \ll 1$, with $\ta = R / \va$ the Alfv\'en travel time and $\va = B/\sqrt{\mu \rho}$ the Alfv\'en velocity. The TB approximation is used here to include the effect of resonant absorption in the inhomogeneous layer, and is valid for $l/R \ll 1$ and $l^\star/R \ll 1$. In the TB approximation, the jump of the perturbations across the inhomogeneous layer is assumed to be the same as their jump across the resonant layer. The expressions for the jump conditions can be found in, e.g., \citet{SGH91,goossens95,tirry} for the static case, and in \citet{goossens92,erdelyi} for the stationary case. Then, the connection formulae at the Alfv\'en resonance  are used as jump conditions for the perturbations at the tube boundary  \citep[see extensive details about the method in][]{goossens06,goossensIAU,goossensSSR}. The dispersion relation for kink and fluting MHD waves in the TT and TB approximations is \citep[see, e.g.,][]{goossens92}
\begin{eqnarray}
 \rhoi \left( \Omega_{\rm i}^2 - \omega_{\rm A i}^2  \right) &+& \rhoe \left( \Omega_{\rm e}^2 - \omega_{\rm A e}^2  \right) = \nonumber \\
&& i \pi \frac{m/ r_{\rm A}}{\rho \left(r_{\rm A} \right) \left| \Delta_{\rm A} \right|} \rhoi \left( \Omega_{\rm i}^2 - \omega_{\rm A i}^2  \right)\rhoe \left( \Omega_{\rm e}^2 - \omega_{\rm A e}^2  \right), \label{eq:reldisper}
\end{eqnarray}
where $\Omega = \omega - U k_z$ is the Doppler-shifted frequency, $\omega_{\rm A}^2 = k_z^2 \va^2$ is the square of the Alfv\'en frequency, $r_{\rm A}$ is the Alfv\'en resonance position, and
\begin{equation}
  \left| \Delta_{\rm A} \right| =  \left| \frac{\rm d}{{\rm d}r} \left[  \Omega^2 - \omega_{\rm A}^2  \right]_{r_{\rm A}}  \right|. \label{eq:delta}
\end{equation}
The term on the right-hand side of Equation~(\ref{eq:reldisper}) contains the effect of resonant absorption. Note that the dependence on $m$ of the dispersion relation is only present in this term. This means that in the TT approximation the effect of the azimuthal wavenumber, $m$, is only felt in the damping of the wave, not in its propagation.

In the case of temporal damping of standing waves, i.e., real $k_z$ and complex $\omega$, the solutions to Equation~(\ref{eq:reldisper}) have been investigated in detail by \citet{goossens92} and \citet{terradasflow}. In the case of spatial damping of propagating waves, i.e., real $\omega$ and complex $k_z$, Equation~(\ref{eq:reldisper}) has been explored by TGV in the absence of flow. Here our purpose is to investigate the effect of flow on the results obtained by TGV in the static case.

 \subsection{No resonant damping}

In the absence of resonant damping, i.e., for $l/R = l^\star/R  = 0$, Equation~(\ref{eq:reldisper}) becomes
\begin{equation}
 \rhoi \left( \Omega_{\rm i}^2 - \omega_{\rm A i}^2  \right) + \rhoe \left( \Omega_{\rm e}^2 - \omega_{\rm A e}^2  \right) = 0. \label{eq:reldisper0}
\end{equation}
Equation~(\ref{eq:reldisper0}) is independent of $m$. In the static case, $\ui = \ue = 0$ and $\Omega_{\rm i} =  \Omega_{\rm e} = \omega$. The solution to Equation~(\ref{eq:reldisper0}) is
\begin{equation}
 k_z = \pm \frac{\omega}{\vk} \equiv \pm k_0, \label{eq:kznoflow}
\end{equation}
with
\begin{equation}
 \vk = \left( \frac{\rhoi \vai^2 + \rhoe \vae^2}{\rhoi + \rhoe} \right)^{1/2},
\end{equation}
the kink velocity. In Equation~(\ref{eq:kznoflow}) the $+$ sign stands for the forward wave and the $-$ sign for the backward wave. In the absence of flow both forward and backward waves are equivalent and have the same wavelength, $\lambda = 2\pi / k_0$.

In the presence of flow, we rewrite Equation~(\ref{eq:reldisper0}) as a second-order polynomial in $k_z$, namely
\begin{equation}
 \left( \vk^2 - \vke^2 \right) k_z^2 + 2 \vcm \omega\, k_z - \omega^2 = 0, \label{eq:poli}
\end{equation}
where $\vcm$ and $\vke$ are defined as
\begin{equation}
 \vcm = \frac{\rhoi \ui + \rhoe \ue}{\rhoi + \rhoe}, \qquad
 \vke =\left( \frac{\rhoi \ui^2 + \rhoe \ue^2}{\rhoi + \rhoe} \right)^{1/2}.
\end{equation}
$\vcm$ is the center-of-mass velocity. $\vke$ is the velocity associated to the kinetic energy of the flow. Note that, similarly, the kink velocity, $\vk$, is the velocity associated to the energy of the magnetic field.  The two solutions to Equation~(\ref{eq:poli}) for $\vke \ne \vk$ are
\begin{equation}
 k_z = - \frac{\vcm}{\vk^2 - \vke^2} \omega \pm k_0 \frac{\vk^2}{\vk^2 - \vke^2} \left[ 1 - \frac{\rhoi \rhoe}{\left( \rhoi + \rhoe \right)^2} \frac{\left( \ui - \ue \right)^2}{\vk^2} \right]^{1/2}, \label{eq:kzsol}
\end{equation}
where the $+$ and $-$ signs in front of the second term stand for forward and backward waves, respectively. If $\ui = \ue = 0$, Equation~(\ref{eq:kzsol}) simply reduces to Equation~(\ref{eq:kznoflow}). We clearly see in Equation~(\ref{eq:kzsol}) that the equivalence between both directions of propagation is broken by the flow.  For $\vke < \vk$ forward and backward waves propagate in opposite directions. For $\vke > \vk$ both waves propagate in the same direction, i.e., they both are forward waves in practice because the flow is strong enough to force the backward wave to reverse its direction of propagation. In the particular case $\vke = \vk$, the solution to Equation~(\ref{eq:poli}) is
\begin{equation}
k_z = \frac{\omega}{2 \vcm},
\end{equation}
which corresponds to the forward wave, while the backward wave does not propagate in the static reference frame. For $\ue = 0$ the condition $\vke = \vk$ is equivalent to
\begin{equation}
 \ui^2 = 2 \vai^2. \label{eq:inversion}
\end{equation}

When the argument of the square root in  Equation~(\ref{eq:kzsol}) is negative, $k_z$ becomes complex. This is the classical KHI \citep[see, e.g.,][]{chandra,drazin}. Then, the two solutions correspond to a spatially damped mode and a spatially amplified mode, respectively. The KHI appears for a critical velocity shear, $\Delta U = \ui - \ue$, defined as
\begin{equation}
 \left( \Delta U\right)^2 > \frac{\left( \rhoi + \rhoe \right)^2}{\rhoi \rhoe} \vk^2 \equiv v_{\rm KH}^2. \label{eq:velKHI}
\end{equation}
Again, in the reference frame where $\ue = 0$ Equation~(\ref{eq:velKHI}) can be rewritten as
\begin{equation}
 \ui^2 > 2 \left( 1 + \frac{\rhoi}{\rhoe} \right) \vai^2. \label{eq:velKHI2}
\end{equation}
From Equations~(\ref{eq:inversion}) and (\ref{eq:velKHI2}) we see that the KHI requires a faster flow velocity than the one needed to reverse the propagation of the backward wave.  

Equivalently to the case without flow (see Equation~(\ref{eq:kznoflow})), we can rewrite Equation~(\ref{eq:kzsol}) as
\begin{equation}
k_z = \frac{\omega}{\vkf} \label{eq:kzgen}
\end{equation}
where we have introduced the effective kink velocity modified by the flow, $\vkf$, defined by
\begin{eqnarray}
\vkf = \left\{  - \frac{\vcm}{\vk^2 - \vke^2} \pm \frac{\vk}{\vk^2 - \vke^2} \left[ 1 - \frac{\rhoi \rhoe}{\left( \rhoi + \rhoe \right)^2} \frac{\left( \ui - \ue \right)^2}{\vk^2} \right]^{1/2} \right\}^{-1}. \nonumber \\ \label{eq:vkf}
\end{eqnarray}
Note that the effective kink velocity is different for forward ($+$ sign) and backward ($-$ sign) waves. This means that the phase speed of the waves depends on their direction of propagation. Also note that $\vkf$ becomes complex beyond the critical velocity shear for the KHI. For slow, sub-Alfv\'enic flows we may drop the quadratic terms in $\ui$ and $\ue$ from Equation~(\ref{eq:vkf}) to obtain a first-order approximation for $\vkf$, namely
\begin{equation}
 \vkf \approx \pm \vk  + \vcm . \label{eq:vkfapp}
\end{equation}
In the absence of flow, Equation~(\ref{eq:vkf}) reduces to $\vkf = \pm \vk$. When the effective kink velocity is equal to the external Alfv\'en velocity, the wave becomes leaky in the external medium. In terms of $k_z$, waves are leaky for wavenumbers larger than
\begin{equation}
k_z = \pm \frac{\omega}{\vae} \equiv \pm k_{\rm L}.
\end{equation}
The forward wave becomes leaky for much slower flow velocities than the backward wave. By using Equation~(\ref{eq:vkfapp}) and taking $\ue = 0$, we find that the forward wave becomes leaky for
\begin{equation}
\ui \gtrsim \left( \frac{\rhoi + \rhoe }{\sqrt{\rhoi \rhoe}} - \sqrt{\frac{2 \left( \rhoi + \rhoe  \right)}{\rhoi}}  \right) \vai. \label{eq:uileaky}
\end{equation}

 Again, we use the approximation of Equation~(\ref{eq:vkfapp}) in Equation~(\ref{eq:kzgen}) to obtain a first-order approximation of the wavelength as
\begin{equation}
 \lambda \approx \lambda_0 \left( \pm 1 + \frac{\vcm}{\vk}  \right), \label{eq:lam}
\end{equation}
with $\lambda_0 = 2\pi / k_0$. We consider the particular case $\ue = 0$ and use the dimensionless notation of TGV to rewrite Equation~(\ref{eq:lam}) as
\begin{equation}
 \frac{\lambda}{R} \approx 2\pi \sqrt{\frac{2\zeta}{\zeta + 1}} \frac{1}{f} \left( \pm 1 + \sqrt{\frac{\zeta}{2 \left( \zeta + 1 \right)}} \bar{\ui}  \right), \label{eq:lam2}
\end{equation}
with
\begin{equation}
 \zeta = \frac{\rhoi}{\rhoe}, \qquad f = \frac{\omega R}{\vai}, \qquad \bar{\ui} = \frac{\ui}{\vai}, \label{eq:dimension}
\end{equation}
the density contrast, the dimensionless frequency, and the dimensionless flow velocity, respectively.

\begin{figure}[!t]
\centering
 \includegraphics[width=0.85\columnwidth]{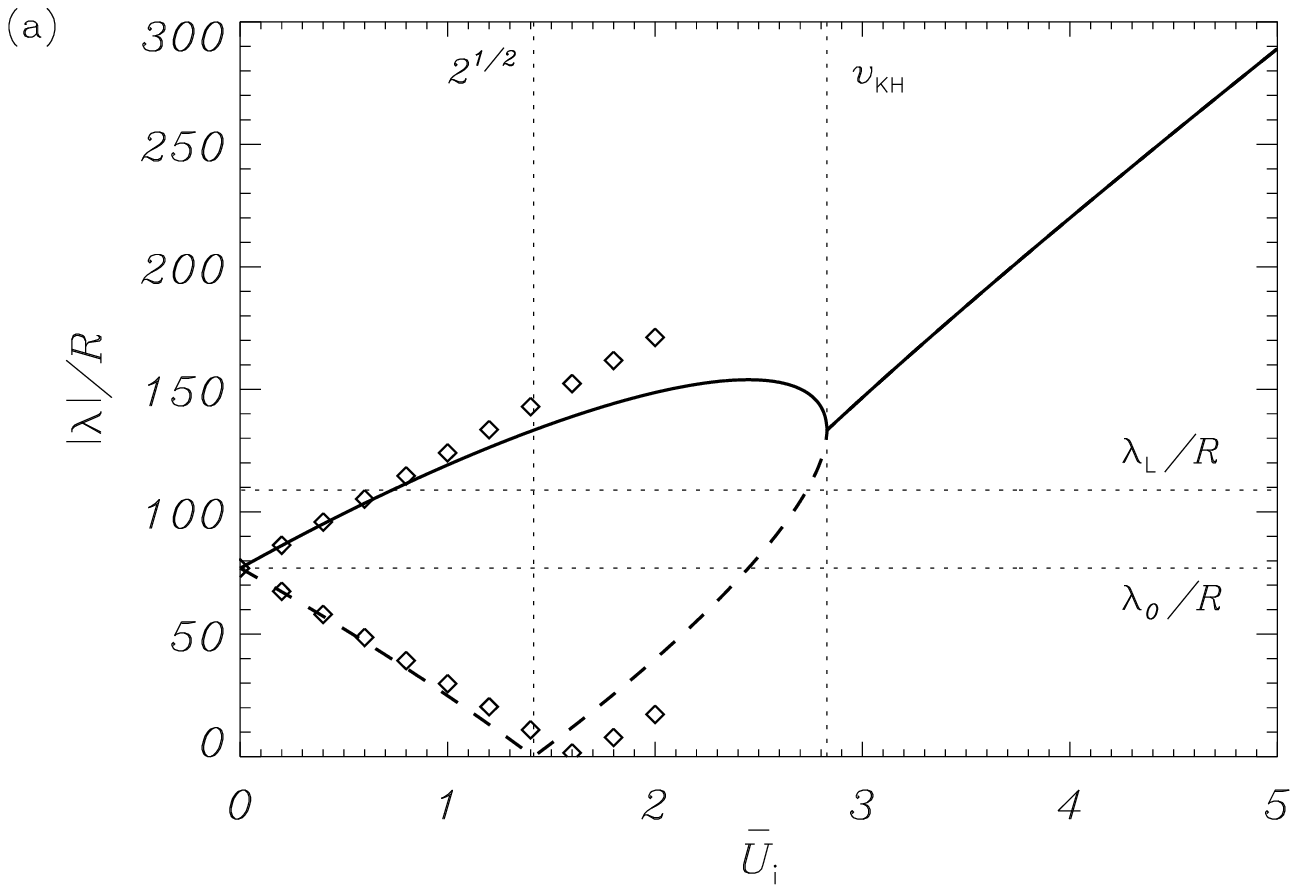}
  \includegraphics[width=0.85\columnwidth]{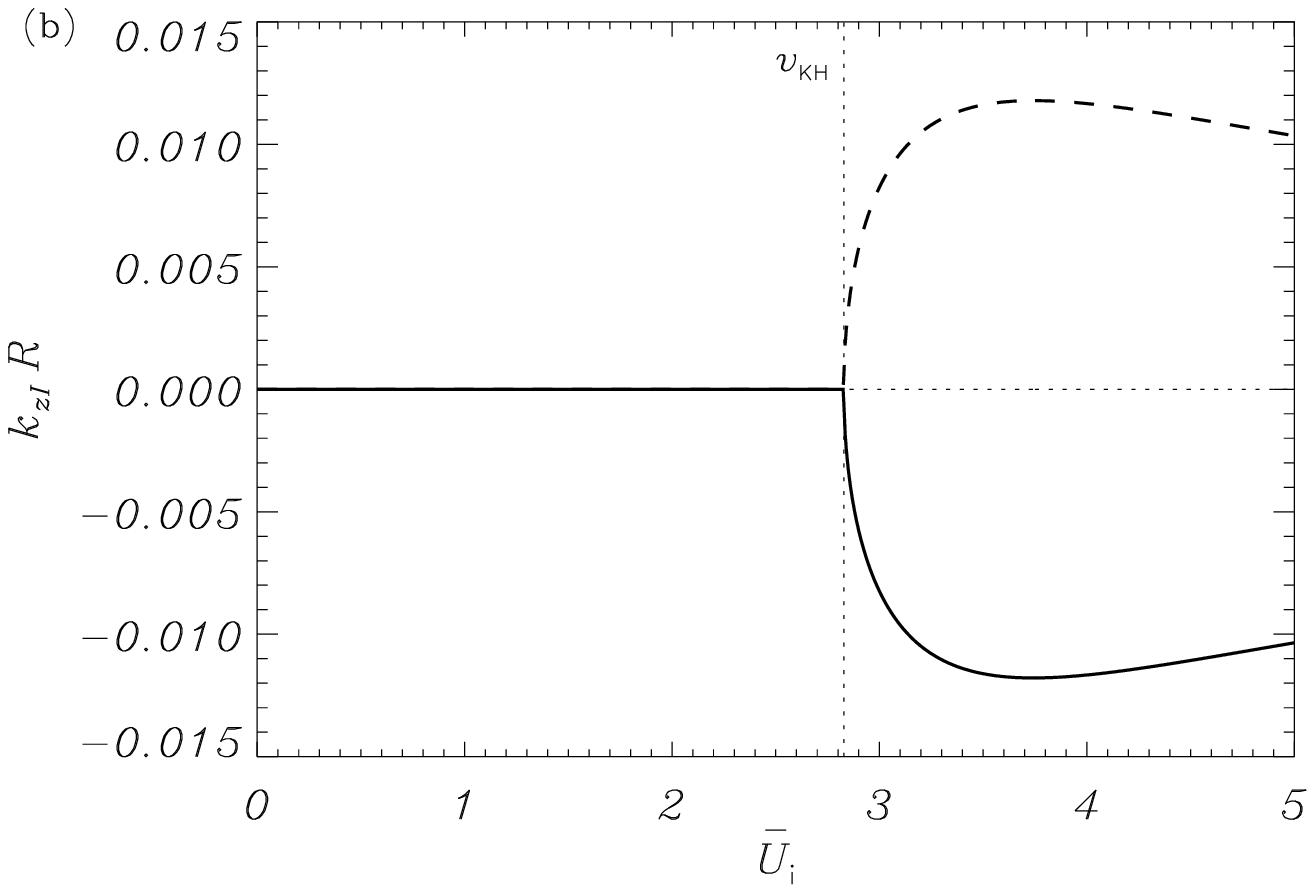}
\caption{(a) $\left| \lambda \right| / R$ and (b) $k_{z \rm I} R$ versus $\bar{\ui}$ corresponding to the forward (solid line) and backward (dashed line) kink waves in the absence of resonant damping. The vertical dotted lines in both panels correspond to the different critical flow velocities indicated in the text. The horizontal dotted line in panel (a) is the wavelength for $\bar{\ui} = 0$, with $\lambda_0 = 2\pi/k_0$. The symbols correspond to the approximation for slow flows given in Equation~(\ref{eq:lam2}). In panel (b) the  horizontal dotted line denotes  $k_{z \rm I} = 0$. In this plot, $f = 0.1$, $\zeta = 3$, and $\ue = 0$. \label{fig:tt}}
\end{figure}

We plot in Figure~\ref{fig:tt}(a) $|\lambda|/R$ versus $\bar{\ui}$ computed from the full Equation~(\ref{eq:kzsol})  for the particular case $\ue = 0$. As predicted by Equation~(\ref{eq:inversion}), the backward wave reverts its direction of propagation for $\bar{\ui} = \sqrt{2}$. For the set of parameters used in Figure~\ref{fig:tt}, the forward wave becomes leaky for a flow velocity slightly sub-Alfv\'enic (Equation~(\ref{eq:uileaky})). When the threshold velocity of the KHI is reached (Equation~(\ref{eq:velKHI2})), both forward and backward waves merge. We compare the full result with the approximation for slow flows given by Equation~(\ref{eq:lam2}) (see the symbols in Fig.~\ref{fig:tt}(a)), and obtain a good agreement for sub-Alfv\'enic flows, i.e., $\bar{\ui} \lesssim 1$. On the other hand, Figure~\ref{fig:tt}(b) displays $k_{z \rm I} R$ versus $\bar{\ui}$. In the absence of resonant damping, the imaginary part of $k_z$ is zero for flow velocities slower than the critical velocity shear of the KHI. For larger velocities, one damped solution and one overstable solution are present.

\subsection{Effect of resonant damping}
\label{sec:resonant}

Here we incorporate the damping due to resonant absorption. We take into account the full expression of the dispersion relation (Equation~(\ref{eq:reldisper})). We write $k_z = k_{z \rm R} + i k_{z \rm I}$ in Equation~(\ref{eq:reldisper}) and consider weak damping, so we neglect terms of $\mathcal{O} \left( k_{z \rm I}^2 \right)$. In addition, we implicitly assume that the flow velocities are slower than the critical velocity of the KHI. After long but straightforward analytical manipulations, we obtain from Equation~(\ref{eq:reldisper}) the expression for the ratio $k_{z \rm I} / k_{z \rm R}$, namely
\begin{equation}
 \left| \frac{k_{z \rm I}}{k_{z \rm R}}\right| = \frac{\pi}{2} \frac{m}{r_{\rm A}} \frac{\rhoi^2}{\rhoi+\rhoe}\frac{1}{\rho \left(r_{\rm A} \right) \left| \Delta_{\rm A} \right|}  \frac{\left( \Omega^2_{\rm i} - \omega_{\rm Ai}  \right)^2}{\omega \left( \omega - \omega_{\rm cm} \right)}, \label{eq:ratio}
\end{equation}
with $\omega_{\rm cm} = k_{z \rm R} \vcm$. In the absence of flows, Equation~(\ref{eq:ratio}) reduces to Equation~(8) of TGV. Due to the different values of $\Omega_{\rm i}$ and $\omega_{\rm cm}$ for forward and backward waves,  Equation~(\ref{eq:ratio}) predicts that both waves have different damping ratios. The expression for the ratio $k_{z \rm I} / k_{z \rm R}$ is more complicated in the presence of flows compared to the static case of TGV. Let us find a more simple expression for $k_{z \rm I} / k_{z \rm R}$ in the limit of slow, sub-Alfv\'enic flows. First, we evaluate $\rho \left(r_{\rm A} \right) \left| \Delta \right|_{\rm A}$ to  explicitly take into account the radial variation of the density and the flow velocity at the resonance position. From Equation~(\ref{eq:delta}) we get
\begin{equation}
 \rho \left(r_{\rm A} \right) \left| \Delta_{\rm A} \right| = \left| \Omega^2 \left(r_{\rm A}\right) \left( \frac{{\rm d} \rho}{{\rm d}r} \right)_{r_{\rm A}} - 2 \rho\left(r_{\rm A}\right) \Omega \left(r_{\rm A}\right)k_{z \rm R} \left( \frac{{\rm d} U}{{\rm d}r} \right)_{r_{\rm A}}   \right|, \label{eq:deltafull}
\end{equation}
where we have used the resonant condition $\Omega^2 \left(r_{\rm A}\right)=  k_{z \rm R}^2 \va^2 \left(r_{\rm A}\right)$. The quantities present in Equation~(\ref{eq:deltafull}) have to be evaluated at the resonance position, $r_{\rm A}$. In the TB approximation it is reasonable to assume $r_{\rm A} \approx R$. There are two terms in the right-hand side of Equation~(\ref{eq:deltafull}). The first term is due to the variation of density and the second term is due to the variation of flow velocity. The direction of wave propagation is also important in Equation~(\ref{eq:deltafull}) because of the sign of $k_{z \rm R}$. In the absence of flow, Equation~(\ref{eq:deltafull}) simplifies to $\rho \left(r_{\rm A} \right) \left| \Delta_{\rm A} \right| =  \omega^2  \left( \frac{{\rm d}\rho}{{\rm d}r}  \right)_{r_{\rm A}}$. 

For smooth profiles, the derivatives of the density and the flow velocity profiles at the resonance position can be cast as
\begin{equation}
\left( \frac{{\rm d}\rho}{{\rm d}r}  \right)_{r_{\rm A}} \approx \mathcal{F} \frac{\pi^2}{4} \frac{\rhoi - \rhoe}{l}, \qquad \left( \frac{{\rm d}U}{{\rm d}r}  \right)_{r_{\rm A}} \approx \mathcal{F} \frac{\pi^2}{4} \frac{\ui - \ue}{l^\star} \label{eq:deriv}
\end{equation}
with $\mathcal{F}$ a factor that depends on the form of the transverse profile. For example, $\mathcal{F} = 4/\pi^2$ for a linear profile \citep{goossens2002} and $\mathcal{F} = 2/\pi$ for a sinusoidal profile \citep{rudermanroberts}. For simplicity we assume the same profile for both density and flow velocity, but we keep $l \ne l^\star$. As values of density and flow velocity at the resonance position we take
\begin{equation}
  \rho \left(r_{\rm A} \right) = \frac{\rhoi + \rhoe}{2}, \qquad  U \left(r_{\rm A} \right) = \frac{\ui + \ue}{2}. \label{eq:valsra}
\end{equation}
We use in Equation~(\ref{eq:deltafull}) the expressions given in Equations~(\ref{eq:deriv}) and (\ref{eq:valsra}) to obtain
\begin{eqnarray}
 \rho \left(r_{\rm A} \right) \left| \Delta_{\rm A} \right| &=& \mathcal{F} \frac{\pi^2}{4} \frac{\rhoi - \rhoe}{l} \Omega^2 \left(r_{\rm A}\right) \nonumber \\
&\times& \left| 1 - \frac{k_{z \rm R}}{\Omega\left(r_{\rm A}\right)} \frac{\rhoi + \rhoe}{\rhoi-\rhoe} \left( \ui - \ue  \right) \frac{l}{l^\star}   \right|. \label{eq:deltafull2}
\end{eqnarray}
For our following analysis, we assume that the second term within the absolute value in Equation~(\ref{eq:deltafull2}) is smaller than one, so that we can drop the absolute value sign. For slow flows, this is equivalent to assume $l^\star \gtrsim l$. In the limit of slow, sub-Alfv\'enic flows, we neglect the quadratic terms in the flow velocities. In addition, we write $k_{z \rm R} \approx \omega / \vkf$ and use the first-order approximation for $\vkf$ given in Equation~(\ref{eq:vkfapp}). Equation~(\ref{eq:deltafull2}) can then be rewritten as
\begin{eqnarray}
 \rho \left(r_{\rm A} \right) \left| \Delta_{\rm A} \right| &\approx& \mathcal{F} \frac{\pi^2}{4} \frac{\rhoi - \rhoe}{l} \omega^2 \nonumber \\
&\times& \left\{ 1 \mp \frac{\ui + \ue }{\vk} \left[ 1 +  \frac{\rhoi + \rhoe}{\rhoi - \rhoe} \frac{\ui - \ue}{\ui + \ue}  \frac{l}{l^\star}    \right]  \right\}.  \label{eq:detafull3}
\end{eqnarray}

Next, we use Equation~(\ref{eq:detafull3}) in Equation~(\ref{eq:ratio}) and perform a first-order expansion in the flow velocities. Equation~(\ref{eq:ratio}) becomes
\begin{eqnarray}
 \left| \frac{k_{z \rm I}}{k_{z \rm R}}\right| &\approx&  \frac{1}{2 \pi} \frac{m}{\mathcal{F}} \frac{l}{R} \frac{ \rhoi-\rhoe }{\rhoi+\rhoe}  \left\{ 1 \pm \frac{\vcm}{\vk} \left[ 1 - 4 \frac{\rhoi + \rhoe}{\rhoi - \rhoe} \frac{ \rhoi \ui - \rhoe \ue}{\rhoi \ui + \rhoe \ue}  \right. \right. \nonumber \\
 &+& \left. \left. \frac{\ui - \ue}{\vcm} \left( 1 + \frac{\rhoi + \rhoe}{\rhoi - \rhoe } \frac{\ui - \ue}{\ui + \ue} \frac{l}{l^\star} \right)  \right] \right\}.  \label{eq:ratioapp}
\end{eqnarray}
As in previous expressions, the $+$ and $-$ signs in Equation~(\ref{eq:ratioapp}) stand for forward and backward waves, respectively. In the particular case $\ue = 0$, Equation~(\ref{eq:ratioapp}) simplifies to
\begin{eqnarray}
\left| \frac{k_{z \rm I}}{k_{z \rm R}}\right| &\approx& \frac{1}{2 \pi} \frac{m}{\mathcal{F}} \frac{l}{R} \frac{ \rhoi-\rhoe }{\rhoi+\rhoe} \nonumber \\
&\times& \left[ 1 \pm  \frac{\ui}{\vk} \left( 1 + \frac{\rhoi}{\rhoi + \rhoe} -  \frac{4\rhoi}{\rhoi - \rhoe} +\frac{\rhoi + \rhoe}{\rhoi - \rhoe }  \frac{l}{l^\star}  \right) \right]. \label{eq:ratiofin}
\end{eqnarray} 
We use again the approximation $k_{z \rm R} \approx \omega / \vkf$ and compute from Equation~(\ref{eq:ratiofin}) the damping length, $\ld = 1/k_{z \rm I}$, as
\begin{equation}
\ld \approx 2\pi \frac{\mathcal{F}}{m} \frac{R}{l} \frac{\rhoi + \rhoe}{\rhoi - \rhoe} \frac{\vk}{\omega} \left[ 1 \pm \frac{\ui}{\vk} \left( \frac{3\rhoi + \rhoe}{\rhoi-\rhoe} - \frac{\rhoi + \rhoe}{\rhoi-\rhoe} \frac{l}{l^\star}  \right) \right]. \label{eq:ld}
\end{equation}
 Finally, we express Equation~(\ref{eq:ld}) using the dimensionless quantities defined in Equation~(\ref{eq:dimension}), namely
\begin{eqnarray}
 \frac{\ld}{R} &\approx& 2\pi \xi_{\rm E} \sqrt{\frac{2\zeta}{\zeta + 1}} \frac{1}{f} \left[ 1 \pm \bar{\ui} \sqrt{\frac{\zeta + 1}{2\zeta}} \left( \frac{3 \zeta + 1}{\zeta - 1} - \frac{\zeta + 1}{\zeta - 1} \frac{l}{l^\star}  \right) \right], \nonumber \\ \label{eq:ldtt}
\end{eqnarray}
with
\begin{eqnarray}
 \xi_{\rm E} = \frac{\mathcal{F}}{m} \frac{R}{l} \frac{\zeta + 1}{\zeta - 1}.
\end{eqnarray}
Equation~(\ref{eq:ldtt}) is the key equation of this investigation and contains basic properties on the spatial damping of propagating kink MHD waves. Several important results can be extracted from Equation~(\ref{eq:ldtt}). Equation~(\ref{eq:ldtt}) predicts that backward and forward propagating waves are damped on length scales that are inversely proportional to the frequency, $f$. This is the same dependence found by TGV in the static case. However, the factor of proportionality depends on the characteristics of the flow and the density contrast, so that the damping length differs from its static analogue. As for the wavelength (see Equation~(\ref{eq:lam2})), the damping length for forward and backward waves is different. 

To shed more light on this result, let us consider the case $l^\star \gg l$. Equation~(\ref{eq:ldtt}) reduces to
\begin{equation}
\ld \approx 2\pi \xi_{\rm E} \sqrt{\frac{2\zeta}{\zeta + 1}} \frac{1}{f} \left( 1 \pm \bar{\ui} \sqrt{\frac{\zeta + 1}{2\zeta}}  \frac{3 \zeta + 1}{\zeta - 1}  \right)  \label{eq:ld2}
\end{equation}
According to Equation~(\ref{eq:ld2}) the backward propagating wave ($-$ sign) gets damped on a shorter length scale than in the absence of flow, while for the forward propagating wave ($+$ sign) the damping length is longer. For $l \approx l^\star$ the damping length for the forward wave remains longer than that of the backward wave, but we must note that for $l^\star \ll l$ the situation may be the opposite. From Equation~(\ref{eq:ldtt}) we can assess the relation between $l$ and $l^\star$ for which the two terms multiplying the flow velocity cancel each other, namely
\begin{equation}
 l^\star = \frac{\zeta + 1}{3\zeta+ 1} l. \label{eq:llstar}
\end{equation}
Thus, the backward wave damping length becomes longer than that of the forward wave for $l^\star$ smaller than the value given in Equation~(\ref{eq:llstar}). Strictly, Equation~(\ref{eq:ldtt}) is not valid in the limit $l^\star \ll l$, because we have to properly take into account the absolute value in Equation~(\ref{eq:deltafull2}). For slow flows and $l^\star \ll l$, Equation~(\ref{eq:deltafull2}) can be approximated as
\begin{equation}
\rho \left(r_{\rm A} \right) \left| \Delta_{\rm A} \right| \approx \mathcal{F} \frac{\pi}{4} \frac{\rhoi + \rhoe}{l^\star} \omega^2 \frac{\ui - \ue}{\vk}. \label{eq:deltalim}
\end{equation}
Importantly, we find that Equation~(\ref{eq:deltalim}) is independent of $l$ and the same expression holds for forward and backward waves. Now, from Equation~(\ref{eq:deltalim}) it is straightforward to obtain that $\ld \sim 1/l^\star$ for both forward and backward waves in the limit $l^\star \ll l$. As discussed by \citet{terradasflow} in the case of temporal damping of standing waves, multiple resonances may occur within the inhomogeneous transitional layer in the limit $l^\star \ll l$ \citep[see Fig.~5 of][]{terradasflow}. In such a case, the total damping rate is the sum of the contributions from each resonance. We do not study this peculiar situation which takes place for very small, probably not realistic $l^\star$. Instead, we refer the reader to \citet{terradasflow} for details.

\section{Numerical computations}
\label{sec:numerics}

Here, we verify the validity of the analytical expressions derived in Section~\ref{sec:analytics} in the TT and TB approximations. To do so, we numerically solve the full eigenvalue problem by means of the PDE2D code \citep{sewell}. The numerical scheme is similar to that used by TGV. The code implements a method based on finite elements to numerically integrate Equations~(\ref{eq:b1}) and (\ref{eq:b2}) in the radial direction from the cylinder axis, $r=0$, to the edge of the numerical domain, $r=r_{\rm max}$. The boundary conditions at $r=0$ are set according the symmetry arguments, while we impose all perturbations to vanish at $r=r_{\rm max}$. In order to obtain a good convergence of the solutions, $r_{\rm max}$ is located far enough from the magnetic tube to avoid numerical errors. We take $r_{\rm max} = 100R$. We use a nonuniform grid with a large density of grid points within the inhomogeneous layer in order to correctly describe the small spatial scales of the eigenfunctions due to the Alfv\'en resonance. To avoid the singularity of the ideal MHD equations at the resonance position, we add to the induction equation (Equation~(\ref{eq:b2})) a resistive term, i.e., $\eta \nabla^2 {\vec b}$, with $\eta$ the magnetic diffusivity. The output of the code is the complex eigenfunctions and their corresponding eigenvalues. In the limit of large $R_{\rm m}$, with $R_{\rm m} =\va R / \eta$ the magnetic Reynolds number, the eigenvalues are independent of $R_{\rm m}$. In such a case, wave damping is due to resonant absorption exclusively. In our computations, we have considered a sufficiently large $R_{\rm m}$ and have checked that the eigenvalues are indeed independent of  $R_{\rm m}$. We typically take $R_{\rm m} \approx 10^7$ in the computations.

The PDE2D code solves the eigenvalue problem for the temporal damping, i.e., for complex $\omega$ provided a fixed and real $k_z$. As in TGV, we need to convert the results from temporal damping to spatial damping. We use Equation~(40) of TGV to perform the conversion from temporal damping (complex $\omega$ and real $k_z$) to spatial damping (real $\omega$ and complex $k_z$). From Equation~(40) of TGV it is straightforward to obtain the relation between the imaginary parts of $\omega$ and $k_z$, namely
\begin{equation}
k_{z \rm I} = \omega_{\rm I} \left( \frac{\partial \omega_{\rm R}}{\partial k_{z \rm R}} \right)^{-1}, \label{eq:relationwk}
\end{equation}
where $\omega_{\rm R}$ and $\omega_{\rm I}$ are the real and imaginary parts of the frequency in the temporal damping case. In the spatial damping case $\omega = \omega_{\rm R}$. The factor within the parenthesis is the group velocity. Note that the flow does not explicitly appear in Equation~(\ref{eq:relationwk}), but its effect is contained in the value of group velocity numerically computed.

First of all, we test the numerical code by considering the case without flow. In this case we fully recover the results of TGV. Thus, we are confident that the code works properly. Hereon we incorporate the effect of flow. Regarding the wavelength, our numerical results indicate that the presence of the transitional layer has a small impact on the value of the wavelength. The wavelength in the case without resonant damping (see Fig.~\ref{fig:tt}(a)) is a very good approximation to the wavelength in the resonant case. The approximation for slow flows given in Equation~(\ref{eq:lam2}) also holds in the case with damping. Therefore, our following analysis is focused on the behavior of $\ld$ in the presence of flows. The numerical results are used to test the approximations behind Equation~(\ref{eq:ldtt}), i.e., the TT and TB approximations, and the assumption of slow flows.

\begin{figure}[!t]
\centering
 \includegraphics[width=0.85\columnwidth]{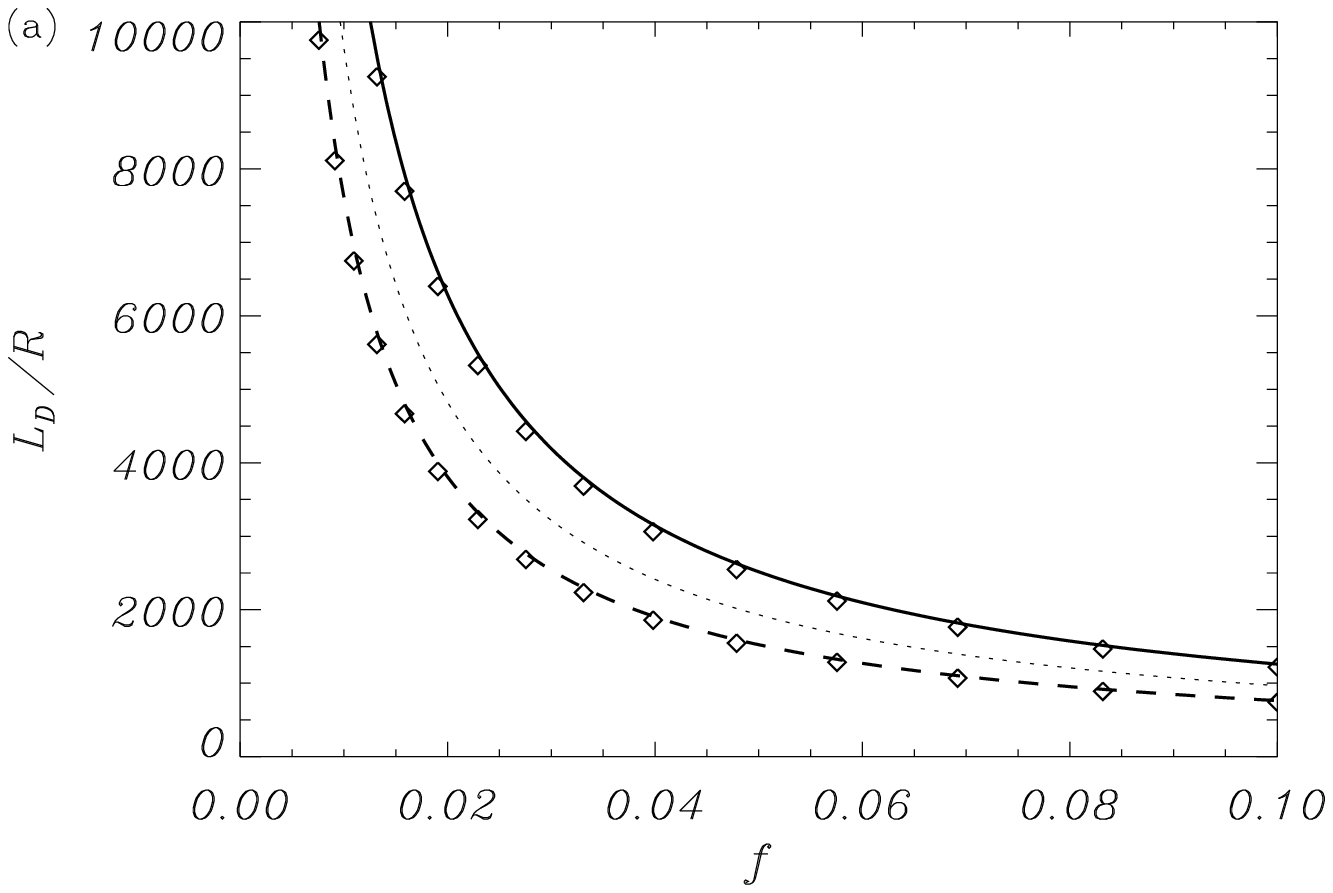}
  \includegraphics[width=0.85\columnwidth]{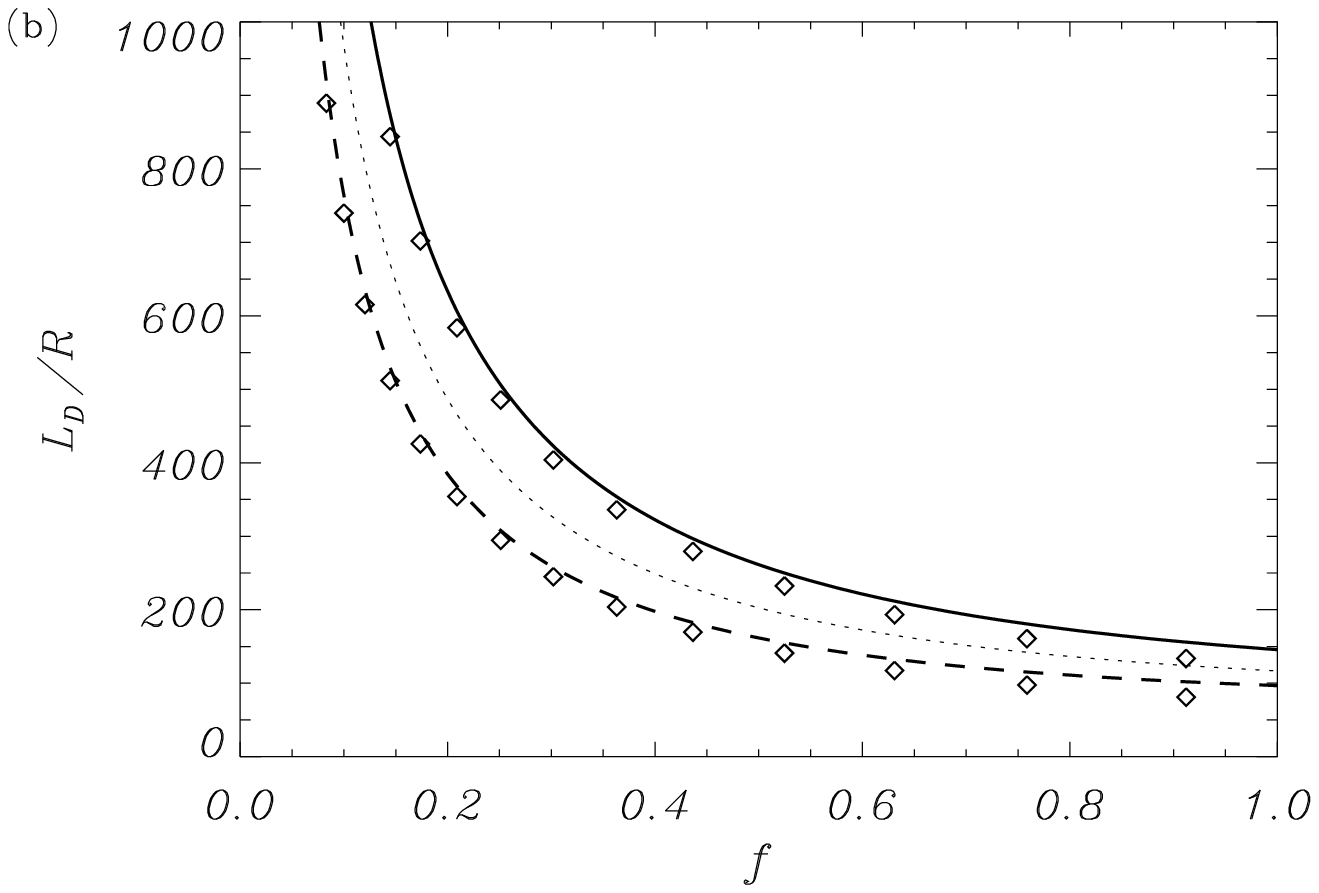}
\caption{(a) Ratio of the damping length to the radius, $\ld / R$,  versus the dimensionless frequency, $f$, corresponding to the forward (solid line) and backward (dashed line) kink waves for $\bar{\ui} = 0.1$, $\ue = 0$, $\zeta = 3$, and $l / R = l^\star / R = 0.1$. The symbols correspond to the approximation for slow flows given in Equation~(\ref{eq:ldtt}). The dotted line is the result in the absence of flow (see TGV). (b) Same as panel (a) but for larger $f$. \label{fig:ld_f}}
\end{figure}

In Figure~\ref{fig:ld_f} we display $\ld/R$ as a function of the dimensionless frequency, $f$, for $\bar{\ui} = 0.1$ (the rest of parameters are indicated in the caption of the Fig.~\ref{fig:ld_f}). As in the static case of TGV, the higher the frequency, the shorter the damping length by resonant absorption.  We obtain the analytically predicted result that forward and backward waves have different damping lengths. In this example, the damping length of the forward wave is longer than that of the backward wave. The equivalent damping length in the absence of flow is in between both values (the dotted line in Fig.~\ref{fig:ld_f}). We compare the numerical results with those in the TT approximation and for slow flows (Equation~(\ref{eq:ldtt})). The TT approximation applies when $f \ll 1$. In Figure~\ref{fig:ld_f}(a) we plot the results for $f \leq 0.1$. An excellent agreement is found between numerical and analytical solutions for both forward and backward waves. The agreement between approximate and numerical results is also remarkably good even when the condition $f \ll 1$ of the TT approximation is not strictly fulfilled. This can be seen in Figure~\ref{fig:ld_f}(b), where results are plotted for $f \leq 1$.

\begin{figure}[!t]
\centering
 \includegraphics[width=0.85\columnwidth]{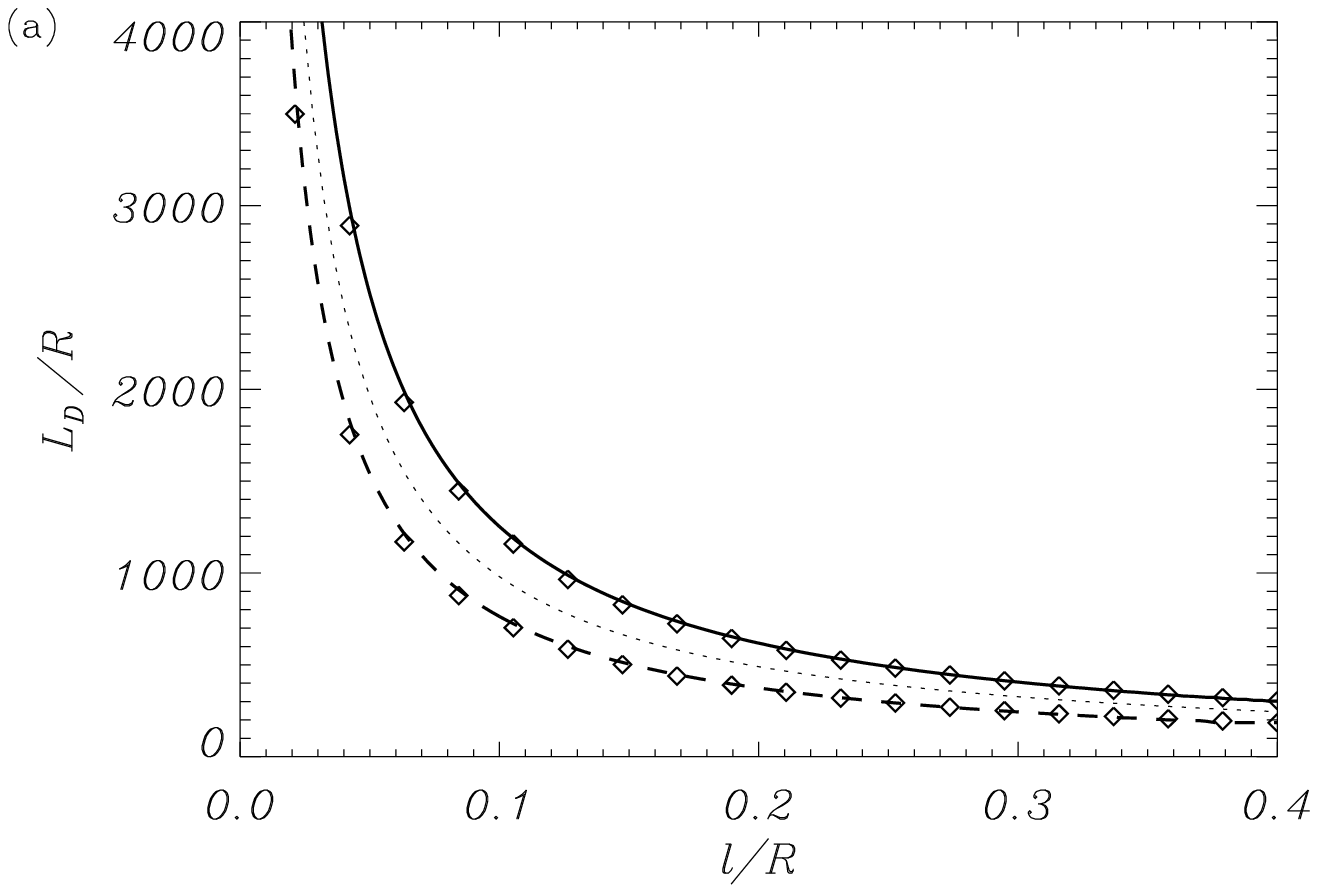}
  \includegraphics[width=0.85\columnwidth]{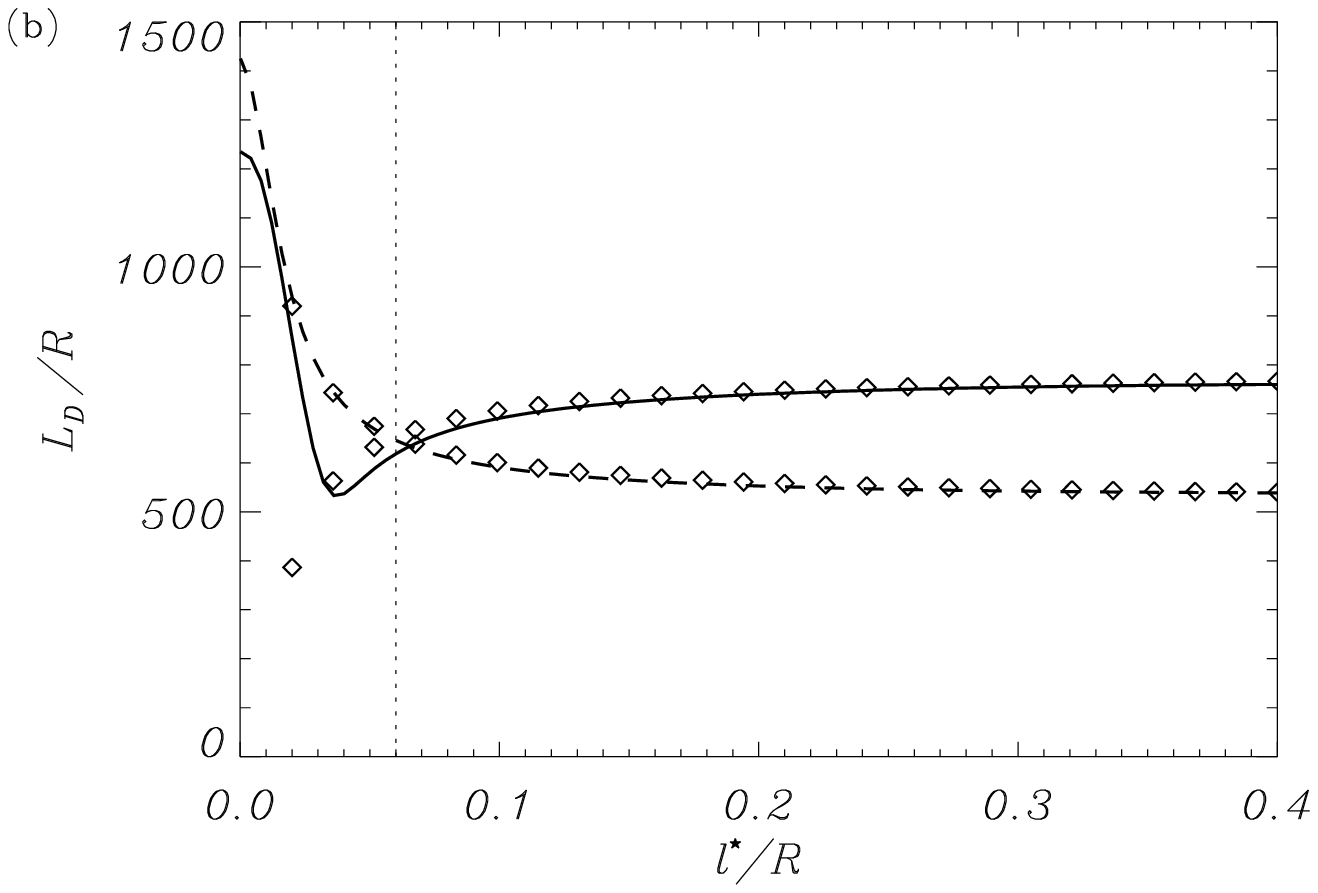}
\caption{(a) Ratio of the damping length to the radius, $\ld / R$, versus $l/R$ for the forward (solid line) and backward (dashed line) kink waves in the case $ l^\star / R = l/R$. The dotted line is the result in the absence of flow. (b) Dependence on $l^\star / R$ for $l/R = 0.15$. The vertical dotted line denotes the value of $l^\star / R$ for which both forward and backward waves have the same $\ld / R$ (Equation~(\ref{eq:llstar})). In both panels the symbols are the approximation of Equation~(\ref{eq:ldtt}). In these plots, $\bar{\ui} = 0.05$, $\ue = 0$, $f = 0.1$, and $\zeta = 3$.\label{fig:ld_l}}
\end{figure}

Next we determine the influence of $l$ and $l^\star$, and test the TB approximation used to derive the analytical expressions. First we consider the case $l = l^\star$. Figure~\ref{fig:ld_l}(a) shows the dependence of the damping length on $l/R$. As for the frequency, $\ld/R$ decreases as $l/R$ increases. The dependence of the damping length on $l/R$ for both forward and backward waves is the same. As before, a very good agreement between Equation~(\ref{eq:ldtt}) and the numerical result is found even when the thickness of the transitional layer departs from the limit $l/R \ll 1$. This means that the TB approximation is sufficiently accurate when the condition $l/R \ll 1$ is slightly relaxed. This result enables us to confidently use Equation~(\ref{eq:ldtt}) beyond the condition of validity of the TB approximation.  Alternately, in Figure~\ref{fig:ld_l}(b) we keep $l/R$ constant and vary $l^\star /R$, so that the spatial scales for the variation of density and flow velocity are different. We recover the analytically predicted result that for $l^\star \gg l$ the results are independent of $l^\star/R$. For $l^\star \ll l$ Equation~(\ref{eq:ldtt}) does not correctly describe the damping length of the forward wave. As discussed at the end of Section~\ref{sec:resonant}, $\ld \sim 1/l^\star$ in the limit $l^\star \ll l$, and so $\ld/R$ increases as $l^\star/R \to 0$.

\begin{figure}[!t]
\centering
 \includegraphics[width=0.85\columnwidth]{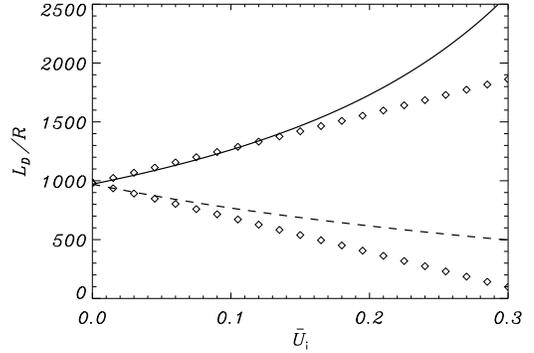}
\caption{Ratio of the damping length to the radius, $\ld / R$,  versus the flow velocity normalized to the internal Alfv\'en velocity, $\bar{\ui}$, for the forward (solid line) and backward (dashed line) kink waves. The symbols are the linear approximation given in Equation~(\ref{eq:ldtt}). We have used $l/R = l^\star /R = 0.1$, $\ue = 0$, $f = 0.1$, and $\zeta = 3$.\label{fig:ld_ui}}
\end{figure}

Finally, we assess the effect of the flow velocity. This is done in Figure~\ref{fig:ld_ui}. Again, the linear approximation (Equation~(\ref{eq:ldtt})) is quite accurate and agrees well with the full numerical results for $\bar{\ui} \lesssim 0.1$. As expected, the difference between the numerical results and the linear approximation increases as the flow velocity gets faster. However, for  $\bar{\ui} = 0.2$ the relative difference between the full solution and the approximation is only around 10\% for the forward wave and 30\% for the backward wave. This means that for realistic flow velocities observed in coronal magnetic loops \citep[e.g.,][]{brekke,winebarger01,winebarger02}, Equation~(\ref{eq:ldtt}) correctly describes the behavior of the damping length.

In summary, in this Section we have confirmed that the analytical expressions obtained in the TT and TB approximations and for slow, sub-Alfv\'enic flows are very accurate even when these expressions are used outside their domain of strict validity. This result enables us to use Equation~(\ref{eq:ldtt}), i.e., the key equation of this investigation, when realistic values of frequency, flow velocity, and the rest of relevant parameters obtained from the observations are used.

\section{Discussion}
\label{sec:discussion}

Naturally, kink waves propagating in nonuniform magnetic flux tubes are spatially damped by resonant absorption. In the static case, TGV showed that the damping length is inversely proportional to the frequency. Here we have investigated analytically and numerically the spatial damping of resonant kink waves in a transversely nonuniform magnetic waveguide in the presence of longitudinal background flow. Longitudinal flow breaks the equivalence between forward and backward propagating waves with respect to the flow direction. The wavelength and the damping length due to resonant absorption are both affected by the flow. For sub-Alfv\'enic flows, the backward wavelength is shorter than that of the forward wave, and backward waves are damped in shorter length scales than forward waves. However, as in TGV we have found that the damping length of both forward and backward propagating waves is inversely proportional to the frequency.

 MHD seismology based on propagating waves has attracted limited attention and definitely less than its counterpart based on standing kink waves. Standing kink MHD waves are rare phenomena as they need a violent and energetic event such as a solar flare for their excitation \citep[see, e.g.,][]{ash,naka}. In the absence of flow and for coronal loop standing oscillations \citep[see, e.g.,][]{nakaofman,goossens2002,arregui07,arregui08,goossens08}, MHD seismology has been used to obtain information of the plasma physical conditions. Particularly, for a given set of parameters provided by the observations, i.e., period, damping time, and wavelength in the case of standing waves, \citet{arregui07} and \citet{goossens08} showed that the possible values of $\vai$, $\zeta$, and $l/R$ which are consistent with the theory form a one-dimensional curve in the three-dimensional parameter space. In principle, any point of this curve can equally explain the observations. \citet{solerfine} and  \citet{arreguiballester} showed that more constrained estimations of $\vai$ and $l/R$ can be given in the case of prominence thread oscillations as the limit $\zeta \gg 1$ can be adopted. More recently, \citet{bayesian} found  that more accurate estimations of the parameters are possible by combining the analytical theory of \citet{goossens08} with statistical Bayesian analysis. In the presence of flows, \citet{terradasletterflow} have recently explained also for standing waves that the flow velocity can be estimated from the wave phase difference along the magnetic loop. 

 On the contrary, propagating MHD waves are ubiquitous in the solar atmosphere \citep[see, e.g.,][]{tomczyk07,tomczyk09} and provide a huge reservoir of possibilities for seismology. Some examples of MHD seismology based on propagating waves are, e.g., \citet{tom08b} using numerical simulations of guided MHD waves by density enhancements in the solar corona, \citet{lin09} using observations of kink waves in prominence threads, \citet{VTG} using resonantly damped kink waves in coronal loops, and \citet{verthspicule} exploiting the properties of kink waves in chromospheric spicules. However, none of these works included flow in their analysis. Our theoretical results given in Equations~(\ref{eq:lam2}) and (\ref{eq:ldtt}) have direct implications for MHD seismology based on propagating waves in a flowing medium, and could be used to infer information about the plasma properties. Therefore, the potential application of MHD seismology to the case of resonantly damped propagating kink waves in a flowing medium must be explored.

In the presence of flow two waves with different  wavelengths and damping lengths but with the same frequency (or period) are simultaneously present. We denote as $\lambda^+$ and ${\ld}^+$ the wavelength and damping length of the forward wave, respectively, and as $\lambda^-$ and ${\ld}^-$ the equivalent quantities of the backward wave. Observationally, this means that it is possible to measure five quantities, namely $\lambda^+$, $\lambda^-$, ${\ld}^+$, ${\ld}^-$, and the period, while the rest of parameters, i.e., $\vai$, $\ui$, $l/R$, and $\zeta$ are in principle unknown. In this analysis we assume $l/R = l^\star /R$ for simplicity. We also take $\lambda^-$ as a positive quantity and so we perform the absolute value of Equation~(\ref{eq:lam2}) when the $-$ sign is used.  If observations can provide us with reliable values for the wavelengths and damping lengths of the two waves, then we can use our theoretical results (Equations~(\ref{eq:lam2}) and (\ref{eq:ldtt})) to obtain seismological estimations for the unknown quantities $\vai$, $\ui$, $l/R$, and $\zeta$ as
\begin{eqnarray}
\va &=& \frac{1}{P} \sqrt{\frac{\zeta + 1}{2 \zeta}} \frac{\lambda^+ + \lambda^-}{2}, \label{eq:seis1} \\
\ui &=& \frac{1}{P} \frac{\zeta + 1}{\zeta} \frac{\lambda^+ - \lambda^-}{2}, \label{eq:seis2}\\
\frac{l}{R} &=& \mathcal{F} \frac{\zeta + 1}{\zeta - 1} \frac{\lambda^+ + \lambda^-}{{\ld}^+ + {\ld}^-}, \label{eq:seis3} \\
\zeta &=& \frac{1 + 2 \gamma}{1 - 2 \gamma}, \label{eq:seis4}
\end{eqnarray}
with $P = 2\pi / \omega$ the period and
\begin{equation}
\gamma = \frac{\lambda^+ - \lambda^-}{\lambda^+ + \lambda^-} \frac{{\ld}^+ + {\ld}^-}{{\ld}^+ - {\ld}^-}.
\end{equation}
In the absence of flow, $\lambda^+ = \lambda^-$ and ${\ld}^+ = {\ld}^-$. Then, Equations~(\ref{eq:seis1}) and (\ref{eq:seis3}) are equivalent to the expressions studied by  \citet{goossens08}, and the density contrast (Equation~(\ref{eq:seis4})) becomes indeterminate. Thanks to the flow, the density contrast can be determined if reliable measures of the wavelengths and damping lengths of both forward and backward waves are available.

The theoretical results of the present paper offer new and exciting opportunities for MHD seismology in plasma structures with equilibrium flows. For the first time we have shown that an estimation of the density contrast, $\zeta = \rhoi/\rhoe$, is possible. MHD seismology requires theory and observations. The required observations might not be available at present time. However, the seismological tool provided in Equations~(\ref{eq:seis1})--(\ref{eq:seis4}) could be used in the future when the required observations become available.

The investigation performed in this paper may be improved in the future by incorporating additional physics in the MHD wave model. Effects that come to mind are the variation of the plasma parameters along the magnetic field direction as in \citet{stratified} and magnetic expansion and twist of the flux tube. These and other effects might be included in forthcoming investigations on propagating resonant kink waves.

\acknowledgements{
  This manuscript was finished during a visit of MG to the Solar Physics Group of UIB. MG is happy to acknowledge the hospitality of the Solar Physics Group and the financial support from UIB through grant 40/2010 under the program ``Estades breus de professors convidats''. We thank I. Arregui for useful comments. RS acknowledges support from a postdoctoral fellowship within the EU Research and Training Network ``SOLAIRE'' (MTRN-CT-2006-035484).  MG acknowledges support from K.U. Leuven via GOA/2009-009. JT  acknowledges support from the Spanish Ministerio de Educaci\'on y Ciencia through a Ram\'on y Cajal grant and funding provided under projects AYA2006-07637 and FEDER funds.}

\end{document}